\documentclass{ws-ijmpa}

\begin{document}

\markboth{A. A. Saharian and E. R. Bezerra de Mello}
{Casimir Densities for Massive Fermionic Fields}

\catchline{}{}{}{}{}
%
%%%%%%%%%%%%%%%%%%%%%%%%%%%%%%%%%%%%%%%%%%%%%%%%%%%%%%%%%%%%%%%%%%%%

\title{CASIMIR DENSITIES FOR A MASSIVE FERMIONIC QUANTUM FIELD IN A GLOBAL
MONOPOLE BACKGROUND WITH SPHERICAL BOUNDARY}

\author{\footnotesize A. A. SAHARIAN}
\address{Department of Physics, Yerevan State University\\
375049 Yerevan, Armenia}
\author{E. R. BEZERRA DE MELLO}
\address{Departamento de F\'{\i}sica-CCEN, Universidade Federal da 
Para\'{\i}ba\\
58.059-970, J. Pessoa, PB C. Postal 5.008, Brazil}

\maketitle

\pub{Received (Day Month Year)}{Revised (Day Month Year)}

\begin{abstract} 
We investigate the vacuum expectation value of the energy-momentum tensor
associated with a massive fermionic field obeying the MIT bag boundary
condition on a spherical shell in the global monopole spacetime. The
asymptotic behavior of the vacuum densities is investigated near the sphere
center and surface, and at large distances from the sphere. In the limit of
strong gravitational field corresponding to small values of the parameter
describing the solid angle deficit in global monopole geometry, the
sphere-induced expectation values are exponentially suppressed.

\end{abstract}

\section{Introduction}

Global monopoles are spherically symmetric topological defects
created due to phase transition when a global symmetry is
spontaneously broken and they have important role in the cosmology
and astrophysics. The simplest theoretical model which provides
global monopoles has been proposed by Barriola and Vilenkin.
\cite{B-V} Neglecting the small size of the monopole's core, the
corresponding metric tensor can be approximately given by the line
element

\begin{equation}
ds^{2}=dt^{2}-dr^{2}-\alpha ^{2}r^{2}\left(d\theta^{2}+\sin^{2}\theta
d\phi^{2}\right) \ ,  \label{mmetric}
\end{equation}
where the parameter $\alpha^{2}$, smaller than unity, depends on
the symmetry breaking energy scale and codifies the presence of
the global monopole. This spacetime corresponds to an idealized
point-like global monopole. Here we shall calculate the vacuum
expectation values of the energy-momentum tensor for fermionic
fields obeying MIT bag boundary condition on the spherical shell
concentric with the global monopole.

\section{Vacuum Energy-Momentum Tensor Inside a Spherical Shell}
\label{sec:inside}

The dynamics of a massive spinor field on a curved spacetime is
described by the Dirac equation
\begin{equation}
i\gamma^{\mu}(\partial_{\mu}+\Gamma_{\mu})\psi-M\psi =0 \ ,
\label{Diraceq}
\end{equation}
where $\gamma^{\mu}$ are the Dirac matrices and $\Gamma_\mu=\frac14
\gamma_{\nu}\nabla_{\mu}\gamma^{\nu}$ is the spin
connection with $\nabla _{\mu }$ being the standard covariant
derivative operator. Our interest in this paper will be the vacuum
expectation values of the energy-momentum tensor induced by a
spherical shell in the global monopole spacetime. We shall assume
that on the sphere surface the field satisfies bag boundary
condition:
\begin{equation}
\left(1+i\gamma^{\mu}n_{b\mu}\right)\psi =0\ ,\quad r=a \ ,
\label{boundcond1}
\end{equation}
where $a$ is the sphere radius, $n_{b\mu }$ is the
outward-pointing normal to the boundary. Expanding the field
operator in terms of the complete set of single-particle states
$\left\{\psi_{\beta}^{(+)},\psi_{\beta }^{(-)}\right\}$ and
making use the standard anticommutation relations, for the vacuum
expectation values of the energy-momentum tensor one finds the
following mode-sum formula
\begin{equation}
\left\langle 0\left\vert T_{\mu \nu }\right\vert 0\right\rangle =\sum_{\beta
}T_{\mu \nu }\left\{ \bar{\psi}_{\beta }^{(-)}(x),\psi _{\beta
}^{(-)}(x)\right\} \ ,  \label{modesum}
\end{equation}
where $|0\rangle $ is the amplitude for the corresponding vacuum.
For the geometry under consideration the eigenfunctions are
specified by the set of quantum numbers $\beta =(kjm\sigma )$,
where  $j=1/2,3/2,\ldots $ determines the value of the total
angular momentum, $m=-j,\ldots ,j$ is its projection, and $\sigma
=0,1$ corresponds to two types of eigenfunctions with different
parities. These functions have the form
\begin{eqnarray}
&&\psi_{\beta}^{(\pm)}=A_\sigma\frac{e^{-i\omega t}}{\sqrt{r}}\left(
\begin{array}{c}
Z_{\nu_{\sigma }}(kr)\Omega_{jlm} \\
in_{\sigma}Z_{\nu _{\sigma}+n_{\sigma}}(kr)\frac{k(\hat{n}\cdot
\vec{\sigma})}{\omega +M}\Omega_{jlm}
\end{array}
\right) \ ,  \quad n_{\sigma }=(-1)^{\sigma } \ , \label{eigfunc} \\
&& \omega=\pm E\ ,\quad E=\sqrt{k^{2}+M^{2}}\ ,\quad \nu_{\sigma}
=\frac{j+1/2}{\alpha }-\frac{n_{\sigma }}{2}
\end{eqnarray}
where $\hat{n}=\vec{r}/r$, $\vec{\sigma}=(\sigma ^{1},\sigma^{2},\sigma^{3})$ 
with the curved space Pauli $2\times 2$ matrices $\sigma ^{k}$. In Eq. 
(\ref{eigfunc}) $Z_{\nu }(x)$ represents a cylindrical function of the order 
$\nu$ and $\Omega_{jlm}$ are the standard spinor spherical harmonics\cite{Berest} 
with $l=j-n_{\sigma }/2$.

For the region inside a spherical shell one has $Z_{\nu}(x)=J_{\nu }(x)$, where 
$J_{\nu }(x)$ is the Bessel function. The imposition of the boundary condition 
on the eigenfunctions (\ref{eigfunc}) leads to the following equations for the
eigenvalues
\begin{equation}
\tilde{J}_{\nu _{\sigma }}(ka)=0\ ,  \label{boundin}
\end{equation}
where for a given function $F(z)$ we use the notation
\begin{equation}
\tilde{F}(z)\equiv zF^{\prime }(z)+\left( \mu +s_{\omega }\sqrt{z^{2}+\mu
^{2}}-(-1)^{\sigma }\nu \right) F(z)\ ,\quad \sigma =0,\ 1,  \label{tildenot}
\end{equation}
with $s_{\omega }={\rm sgn}(\omega)$ and $\mu=Ma$. Let us denote by 
$\lambda_{\nu _{\sigma },s}=ka$, $s=1,2,\ldots ,$ the roots to
equation (\ref{boundin}) in the right half plane, arranged in
ascending order. By using the standard integral for the Bessel
functions, for the normalization coefficient one finds
\begin{equation}
A_{\sigma }^{2}=\frac{z}{2\alpha ^{2}a^{2}}\frac{\sqrt{z^{2}+a^{2}M^{2}}+aM}{
\sqrt{z^{2}+a^{2}M^{2}}}T_{\nu _{\sigma }}(z)\ ,\ \ z=\lambda_{\nu_{\sigma},s}
\ \ ,  \label{normcoef}
\end{equation}
with
\begin{equation}
T_{\nu }^{-1}(z)=\frac{J_{\nu }^{2}(z)}{z}\left[z^{2}+(\mu-(-1)^{\sigma}\nu)
(\mu +s_{\omega }\sqrt{z^{2}+\mu ^{2}})-\frac{s_{\omega }z^{2}}{2\sqrt{z^{2}+
\mu ^{2}}}\right] \ .  \label{r1}
\end{equation}

Because the spacetime is spherically symmetric and static it follows that the vacuum
energy-momentum tensor is diagonal:
\begin{equation}
\left\langle 0\left\vert T_{\mu}^{\nu}\right\vert 0\right\rangle={\rm diag
}(\varepsilon ,-p,-p_{\perp },-p_{\perp })\ ,  \label{diagform}
\end{equation}
with the energy density $\varepsilon$, radial, $p$, and azimuthal, $p_{\perp}$, 
pressures being functions on the radial coordinate only. As a consequence of the 
continuity equation $\nabla_{\nu }\left\langle 0\left\vert 
T_{\mu}^{\nu}\right\vert 0\right\rangle =0$, these functions are related by the 
equation
\begin{equation}
r\frac{dp}{dr}+2(p-p_{\perp })=0\ ,  \label{conteq}
\end{equation}
which means that the radial dependence of the radial pressure necessarily
leads to the anisotropy in the vacuum stresses. Below we will give formulas
for $\varepsilon $ and $p$. The azimuthal pressure is expressed via the
radial one by using formula (\ref{conteq}).

Substituting eigenfunctions (\ref{eigfunc}) into Eq. (\ref{modesum}), the
summation over the quantum number $m$ can be done by using standard
summation formula for the spherical harmonics. For the energy-momentum
tensor components one finds
\begin{eqnarray}
q(r)&=&\frac{-1}{8\pi\alpha^{2}a^{3}r}\sum_{j=1/2}^{\infty}(2j+1)
\sum_{\sigma=0,1}\sum_{s=1}^{\infty }T_{\nu_{\sigma }}(\lambda_{\nu_{\sigma },s})
f_{\sigma \nu _{\sigma }}^{(q)}\left[\lambda_{\nu_{\sigma },s},
J_{\nu_{\sigma}}(\lambda_{\nu _{\sigma },s}r/a)\right] \ , \nonumber\\
q&=&\varepsilon ,\ p,\   \label{qrin}
\end{eqnarray}
where we have introduced the notations
\begin{eqnarray}
f_{\sigma \nu }^{(\varepsilon )}\left[ z,J_{\nu }(y)\right]  &=&z\left[ (
\sqrt{z^{2}+\mu ^{2}}-\mu )J_{\nu }^{2}(y)+(\sqrt{z^{2}+\mu ^{2}}+\mu
)J_{\nu +n_{\sigma }}^{2}(y)\right] \ ,  \label{fnueps} \\
f_{\sigma \nu }^{(p)}\left[ z,J_{\nu }(y)\right]  &=&\frac{z^{3}}{\sqrt{
z^{2}+\mu ^{2}}}\left[ J_{\nu }^{2}(y)-\frac{2\nu +n_{\sigma }}{y}J_{\nu
}(y)J_{\nu +n_{\sigma }}(y)+J_{\nu +n_{\sigma }}^{2}(y)\right] \ .
\label{fnupperp}
\end{eqnarray}

Applying to the sums over $s$ the generalized Abel-Plana summation 
formula\cite{Saha00}, the components of the vacuum energy-momentum tensor can be
presented in the form
\begin{equation}
q(r)=q_{m}(r)+q_{b}(r),\quad q=\varepsilon ,p,p_{\perp }\ ,  \label{qm+qb}
\end{equation}
where $q_{m}(r)$ does not depend on the radius of the sphere $a$ and is the
contribution due to unbounded global monopole spacetime. The corresponding
quantities for the massless case are investigated in Ref. \refcite{EVN}. 
The second term on the right of formula (\ref{qm+qb}) is induced by the presence 
of the spherical shell and can be presented in the form
\begin{equation}
q_{b}(r)=\frac{1}{\pi ^{2}\alpha ^{2}a^{3}r}\sum_{l=1}^{\infty }l\int_{\mu
}^{\infty }\frac{x^{3}dx}{\sqrt{x^{2}-\mu ^{2}}}\frac{W\left[ I_{\nu
}(x),K_{\nu }(x)\right] }{W\left[ I_{\nu }(x),I_{\nu }(x)\right] }F_{\nu
}^{(q)}\left[ x,I_{\nu _{\sigma }}(xr/a)\right] \ ,  \label{qb}
\end{equation}
with
\begin{eqnarray}
F_{\nu }^{(\varepsilon )}\left[ x,I_{\nu }(y)\right]  &=&\left( 1-\frac{\mu
^{2}}{x^{2}}\right) \left\{ I_{\nu -1}^{2}(y)-I_{\nu }^{2}(y)-\mu \frac{
I_{\nu -1}^{2}(y)+I_{\nu }^{2}(y)}{W\left[ I_{\nu }(x),K_{\nu }(x)\right]}
\right\} \ ,  \label{Fnueps} \\
F_{\nu }^{(p)}\left[ x,I_{\nu }(y)\right]  &=&I_{\nu -1}^{2}(y)-I_{\nu
}^{2}(y)-\frac{2\nu -1}{y}I_{\nu }(y)I_{\nu -1}(y)\ .  \label{Fnupperp}
\end{eqnarray}
Here and below $l=j+1/2$, $\nu \equiv \nu _{1}=l/\alpha +1/2$, and
for given functions $f(x)$ and $g(x)$ we use the notation
\begin{eqnarray}
W\left[ f(x),g(x)\right]&=&\left[ xf^{\prime }(x)+(\mu +\nu )f(x)\right] 
\left[ xg^{\prime }(x)+(\mu +\nu )g(x)\right]\nonumber\\
&+&(x^{2}-\mu ^{2})f(x)g(x).
\label{Wnot}
\end{eqnarray}

It can be easily checked that for a massless spinor field the
boundary-induced part of the vacuum energy-momentum tensor is traceless and
the trace anomalies are contained only in the purely global monopole part
without boundaries.

Now we turn to the consideration of various limiting cases of the
expressions for the sphere-induced vacuum expectation values. In the limit $
r\rightarrow 0$, for the boundary parts (\ref{qb}) the summands with a given
$l$ behave as $r^{2l/\alpha -2}$, and the leading contributions come from
the lowest $l=1$ terms. Making use standard formulae for the Bessel modified
functions for small values of the argument, for the sphere-induced parts
near the center, $r\ll a$, one finds
\begin{eqnarray}
\varepsilon _{b}&\approx &\frac{\pi ^{-2}a^{-4}\left( r/2a\right) ^{\frac{2
}{\alpha }-2}}{2\alpha ^{2}\Gamma ^{2}\left( \frac{1}{\alpha }+\frac{1}{2}
\right) }\int_{\mu }^{\infty }dx\,x^{\frac{2}{\alpha }}\sqrt{x^{2}-\mu ^{2}}
\frac{W\left[ I_{\nu }(x),K_{\nu }(x)\right] -\mu }{W\left[ I_{\nu
}(x),I_{\nu }(x)\right] } \ ,  \label{epsbrto0} \\
p_{b} &\approx &\frac{\pi ^{-2}a^{-4}\left( r/2a\right) ^{\frac{2}{\alpha }
-2}}{2\alpha (2+\alpha )\Gamma ^{2}\left( \frac{1}{\alpha }+\frac{1}{2}
\right) }\int_{\mu }^{\infty }\frac{x^{\frac{2}{\alpha }+2}dx}{\sqrt{
x^{2}-\mu ^{2}}}\frac{W\left[ I_{\nu }(x),K_{\nu }(x)\right] }{W\left[
I_{\nu }(x),I_{\nu }(x)\right] } \ ,  \label{pbrto0}
\end{eqnarray}
where $\nu =1/\alpha+1/2$ and $\Gamma(x)$ is the gamma function. Hence, at
the sphere center the boundary parts vanish for the global monopole
spacetime ($\alpha <1$) and are finite for the Minkowski spacetime ($\alpha
=1$). Note that in the large mass limit, $\mu\gg 1$, the integrals in Eqs. (
\ref{epsbrto0}), (\ref{pbrto0}) are exponentially suppressed by the factor $
e^{-2\mu }$.

\begin{figure}[tbph]
\begin{center}
\begin{tabular}{cc}
\epsfig{figure=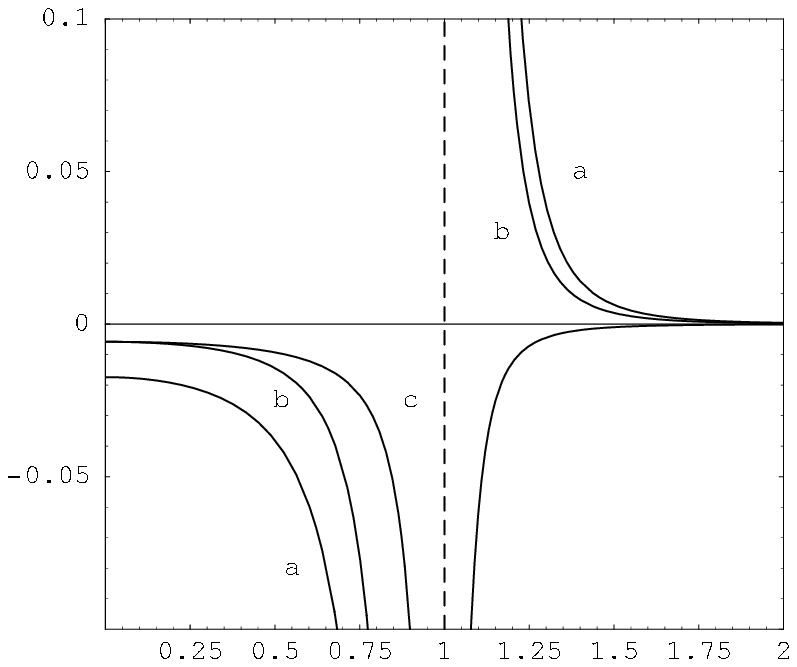,width=5.5cm,height=5cm} & \quad 
\epsfig{figure=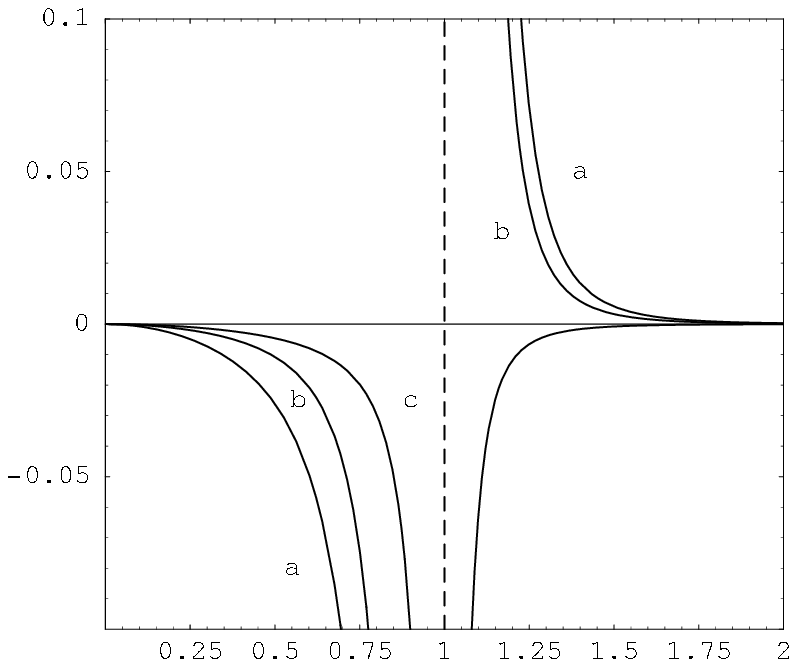,width=5.5cm,height=5cm}
\end{tabular}
\end{center}
\caption{Vacuum energy density, $a^{4}\protect\varepsilon $ (curve
a), azimuthal pressure $a^{4}p_{\perp }$ (curve b), and radial
pressure $a^{4}p$ (curve c) for a massless spinor as functions on
the ratio $r/a$ inside and outside a spherical shell in the
Minkowski spacetime ($\protect\alpha =1$, left panel) and for the
global monopole spacetime with $\protect\alpha =0.5$ (right
panel).} \label{fig1Mink}
\end{figure}

The boundary induced parts of the vacuum energy-momentum tensor
components diverge at the sphere surface, $r\rightarrow a$. In
order to find the leading terms of the corresponding asymptotic
expansion in powers of the distance from the sphere surface, we
note that in the limit $r\rightarrow a$ the sum over $l$ in
(\ref{qb}) diverges and, hence, for small $1-r/a$ the main
contribution comes from the large values of $l$. Consequently,
rescaling the integration variable $x\rightarrow \nu x$ and making
use the uniform asymptotic expansions for the modified Bessel
functions for large values of the order, to the leading order one
finds
\begin{equation}
\varepsilon _{b}(r)\approx -\frac{\mu +1/5}{12\pi ^{2}a(a-r)^{3}}\ ,\quad
p_{b}(r)\approx -\frac{1/5-2\mu }{24\pi ^{2}a^{2}(a-r)^{2}}\ .  \label{pbas}
\end{equation}
Notice that the terms in these expansions diverging as the inverse
fourth power of the distance have canceled out. This is a
consequence of the conformal invariance of the massless fermionic
field. It is of interest to note that the leading terms do not
depend on the parameter $\alpha $ and, hence, are the same for the
Minkowski and global monopole bulks. In Fig. \ref{fig1Mink} we
have presented the dependence of the Casimir densities,
$a^{4}q_{b}(r)$ on the rescaled radial coordinate $r/a$ for a
massless spinor field on the Minkowski bulk (left panel) and on
the global monopole background with the solid angle deficit
parameter $\alpha =0.5$ (right panel). The vacuum energy density
and pressures are negative inside the sphere.

\begin{figure}[tbph]
\begin{center}
\begin{tabular}{cc}
\epsfig{figure=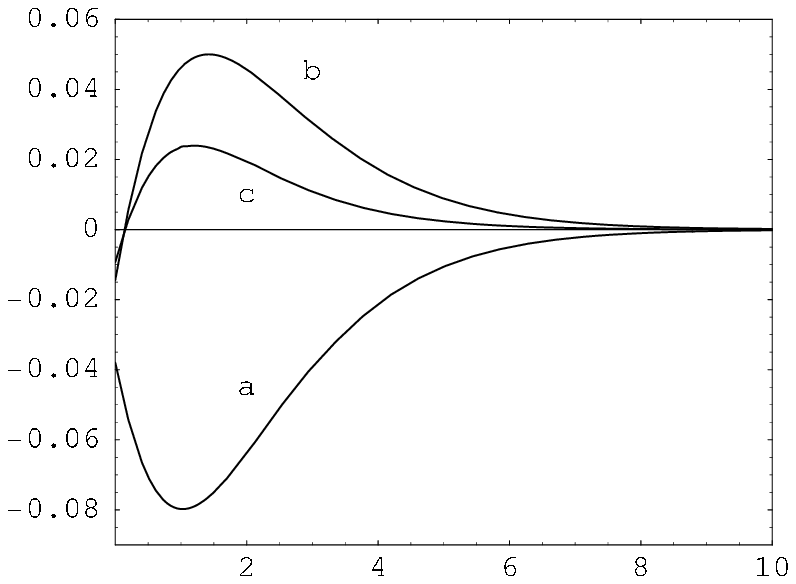,width=5.5cm,height=5cm} & \quad 
\epsfig{figure=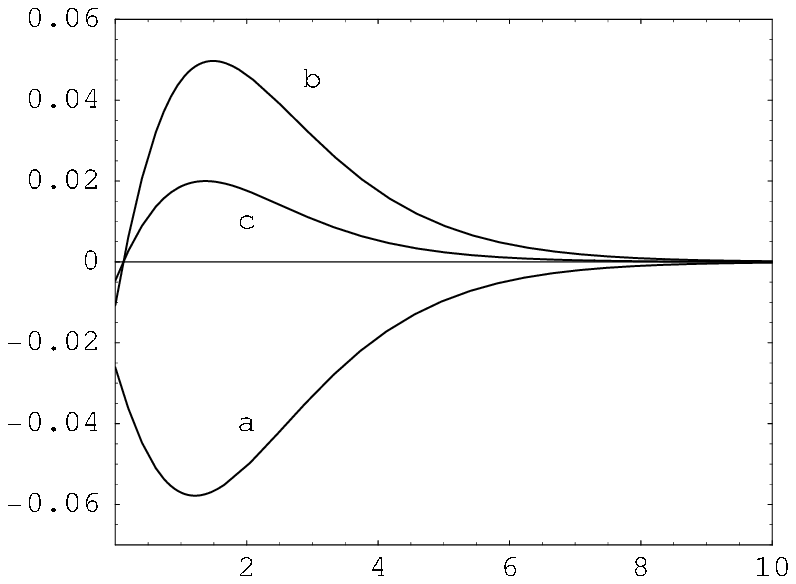,width=5.5cm,height=5cm}
\end{tabular}
\end{center}

\caption{Boundary-induced vacuum expectation values $a^{4}q_{b}(r)$, $q=
\protect\varepsilon ,p,p_{\perp }$, as functions on $\protect\mu =Ma$ for $
r/a=0.5$. The curves a, b, c correspond to the energy density ($\protect
\varepsilon $), azimuthal pressure ($p_{\perp }$), and radial pressure ($p$
), respectively. For the left panel $\protect\alpha =1$ (Minkowski
spacetime) and for the right panel $\protect\alpha =0.5$.}
\label{fig3mass}
\end{figure}

Now let us consider the limit $\alpha \ll 1$ for a fixed value
$r<a$. This limit corresponds to strong gravitational fields. In
this case one has $\nu \approx l/\alpha \gg 1$. The main
contribution to the sum over $l$ comes from the summands with
$l=1$ and the boundary parts of the vacuum energy-momentum tensor
components behave as $\exp [-2\ln (a/r)/\alpha ]$ with
$p_{b}/p_{\perp b}\sim \alpha $. Hence, for $\alpha \ll 1$ the
boundary-induced vacuum expectation values are exponentially
suppressed and the corresponding vacuum stresses are strongly
anisotropic. Fig. \ref{fig3mass} shows that the nonzero mass can
essentially change the behavior of the vacuum energy-momentum
tensor components. In this figure we
have depicted the dependence of the boundary induced quantities $%
a^{4}q_{b}(r)$ on the parameter $Ma$ for the radial coordinate $r=0.5a$. The
left panel corresponds to the sphere in the Minkowski spacetime ($\alpha =1$%
) and for the right panel $\alpha =0.5$.

\section{Vacuum Expectation Values Outside a Spherical Shell}
\label{sec:outside}

Now let us consider the expectation values of the energy--momentum
tensor in the region outside a spherical shell, $r>a$. The
corresponding eigenfunctions have the form (\ref{eigfunc}) with
\begin{equation}
Z_{\nu }(kr)=g_{\nu }(ka,kr)\equiv J_{\nu }(kr)\tilde{Y}_{\nu }(ka)-Y_{\nu
}(kr)\tilde{J}_{\nu }(ka)\ ,  \label{gnu}
\end{equation}
where $Y_{\nu }(z)$ is the Neumann function, and the functions
with tilda are defined as (\ref{tildenot}). Now the spectrum for
the quantum number $k$ is continuous and for the normalization
coefficient we obtain
\begin{equation}
A_{\sigma }^{2}=\frac{k(\omega +M)}{2\alpha ^{2}\omega \lbrack \tilde{J}
_{\nu _{\sigma }}^{2}(ka)+\tilde{Y}_{\nu _{\sigma }}^{2}(ka)]}\ .
\label{Aout}
\end{equation}
Substituting the eigenfunctions (\ref{eigfunc}) into the mode-sum formula 
(\ref{modesum}) and taking into account Eqs. (\ref{gnu}) and (\ref{Aout}), we
can see that the vacuum energy-momentum tensor has the form (\ref{diagform}). 
The diagonal components are determined by formulae
\begin{equation}
q(r)=\frac{-1}{8\pi \alpha ^{2}a^{3}r}\sum_{j=1/2}^{\infty
}(2j+1)\sum_{\sigma =0,1}\int_{0}^{\infty }dx\frac{f_{\sigma \nu _{\sigma
}}^{(q)}\left[ x,g_{\nu _{\sigma }}(x,xr/a)\right] }{\tilde{J}_{\nu _{\sigma
}}^{2}(x)+\tilde{Y}_{\nu _{\sigma }}^{2}(x)}\ ,\quad q=\varepsilon ,\ p,\
\label{qrout}
\end{equation}
where the expressions for $f_{\sigma \nu _{\sigma }}^{(q)}\left[
x,g_{\nu _{\sigma }}(x,xr/a)\right] $ are obtained from formulae
(\ref{fnueps}), (\ref{fnupperp}) by replacements
\begin{equation}
J_{\nu }(y)\rightarrow g_{\nu }(x,y)\ ,\quad J_{\nu +n_{\sigma
}}(y)\rightarrow J_{\nu +n_{\sigma }}(y)\tilde{Y}_{\nu }(ka)-Y_{\nu
+n_{\sigma }}(y)\tilde{J}_{\nu }(ka)\ .  \label{Jtog}
\end{equation}
To find the parts in the vacuum expectation values of the energy-momentum
tensor induced by the presence of the sphere we subtract the corresponding
components for the monopole bulk without boundaries. After rotation of the
integration contour and introducing the Bessel modified functions, for the
boundary-induced parts one obtains
\begin{equation}
q_{b}(r)=\frac{1}{\pi ^{2}\alpha ^{2}a^{3}r}\sum_{l=1}^{\infty }l\int_{\mu
}^{\infty }\frac{x^{3}dx}{\sqrt{x^{2}-\mu ^{2}}}\frac{W\left[ I_{\nu
}(x),K_{\nu }(x)\right] }{W\left[ K_{\nu }(x),K_{\nu }(x)\right] }F_{\nu
}^{(q)}\left[ x,K_{\nu }(xr/a)\right] .  \label{qbout}
\end{equation}
Here the expressions for the functions $F_{\sigma\nu}^{(q)}\left[x,
K_{\nu}(y)\right] $ are obtained from formulae
(\ref{Fnueps}), (\ref{Fnupperp}) by replacements $I_{\nu
}(y)\rightarrow K_{\nu }(y)$ and $I_{\nu -1}(y)\rightarrow -K_{\nu
-1}(y)$. As for the interior region, it can be seen that in the
limit of strong gravitational fields, $\alpha \ll 1$, the boundary
induced vacuum expectation values are exponentially suppressed by
the factor $\exp [-(2/\alpha )\ln (r/a)]$, an the corresponding
vacuum stresses are strongly anisotropic: $p_{b}/p_{\perp b}\sim\alpha $.

In Fig. \ref{fig1Mink} we have plotted the dependence of the vacuum energy
density and stresses on the radial coordinate for a massless spinor field
outside a sphere on the Minkowski bulk (left panel) and on background of the
global monopole spacetime with $\alpha =0.5$. As seen form these figures,
the energy density and azimuthal pressure are positive outside a sphere, and
the radial pressure is negative. The latter has the same sign as for the
interior region.

For the case of a massless spinor the asymptotic behavior of boundary part 
(\ref{qbout}) at large distances from the sphere can be obtained by
introducing a new integration variable $y=xr/a$ and expanding the
subintegrands in terms of $a/r$. The leading contribution for the
summands with a given $l$ has an order $(a/r)^{2\nu +4}$ and the
main contribution comes from the $l=1$ term. The leading terms for
the asymptotic expansions over $a/r$ can be presented in the form
\begin{equation}  \label{qblarger}
q_b(r)\approx \frac{1}{2^{\frac{2}{\alpha }}\pi a^4}\frac{\Gamma \left(\frac{
1}{\alpha }+1\right) \Gamma\left( \frac{2}{\alpha } +\frac{3}{2}\right) f_q}{
(4-\alpha ^2)(2+\alpha )\Gamma ^3\left( \frac{1}{\alpha } +\frac{1}{2}
\right) } \left( \frac{a}{r}\right) ^{\frac{2}{\alpha }+5} \ ,
\end{equation}
where
\begin{equation}  \label{fqlarge}
f_{\varepsilon }=4\frac{\alpha +1}{3\alpha +2} \ ,\quad f_p=-\frac{2\alpha }
{3\alpha+2} \ , \quad f_{p_{\perp }}=1 \ .
\end{equation}

As for the interior components, the quantities (\ref{qbout}) diverge at the
sphere surface $r=a$. Near the surface the dominant contributions come from
modes with large $l$ and by making use the uniform asymptotic expansions for
the Bessel modified functions, the asymptotic expansions can be derived in
powers of the distance from the sphere. The leading terms of these
asymptotic expansions are determined by formulae
\begin{equation}
\varepsilon _{b}(r)\sim \frac{1/5-5\mu }{12\pi ^{2}a(r-a)^{3}}\ ,\quad
p_{b}(r)\sim -\frac{1/5-2\mu }{24\pi ^{2}a^{2}(r-a)^{2}}\ .  \label{pbasout}
\end{equation}

Recall that near the sphere the interior energy density is always negative.
As we see, the leading terms for the radial pressure are the same for the
regions outside and inside the sphere. For the azimuthal pressure these
terms have opposite signs. In the case of the massless spinor field the same
is true for the energy density.

\section{Concluding Remarks}
\label{sec:conc}

In this paper we have analyzed the fermionic Casimir densities
induced by a spherical shell in a idealized point-like global
monopole spacetime. The boundary-induced expectation values for
the components of the energy-momentum tensor are given by formulae (\ref{qb}
) and (\ref{qbout}) for interior and exterior regions, respectively. These
expressions diverge in a non-integrable manner as the boundary is
approached. The energy density and azimuthal pressure vary, to leading
order, as the inverse cube of the distance from the sphere, and near the
sphere the azimuthal pressure has opposite signs for the interior and
exterior regions. For a massless spinor the same is true for the energy
density. The radial pressure varies as the inverse square of the distance
and near the sphere has the same sign for exterior and interior regions.
The leading terms of the corresponding asymptotic expansions near the sphere
do not depend on the solid angle deficit parameter and are the same for
these two cases. Near the sphere the interior energy density is negative for
all values of the mass, while the exterior energy density is positive for $
Ma<0.04$ and is negative for $Ma>0.04$. Near the sphere center the
dominant contributions come from modes with $l=1$ and the
sphere-induced vacuum expectation values vanish for the global
monopole spacetime and are finite for the Minkowski bulk. At large
distances from the sphere the components of the vacuum
energy-momentum tensor go to zero as $(a/r)^{2/\alpha +5}$. In the
limit of strong gravitational field, corresponding to small values
of the parameter $\alpha $, describing the solid angle deficit,
the boundary-induced part of the vacuum energy-momentum tensor is
strongly suppressed by the factor $\exp [-(2/\alpha )|\ln (r/a)|]$
and the corresponding vacuum stresses are strongly anisotropic:
$p_{b}\sim \alpha p_{\perp b}$. Note that this suppression effect
also takes place in the scalar case. \cite{A-M}

\end{document}